\documentclass{PoS}

\title{The GoSam package: an overview}

\ShortTitle{The GoSam package: an overview}

\author{\speaker{Francesco Tramontano}%
        \thanks{On behalf of the GOSAM collaboration: 
G.~Cullen, N.~Greiner, G.~Heinrich, G.~Luisoni, P.~Mastrolia, G.~Ossola,
T.~Reiter, F.~Tramontano, H.~van~Deurzen, J.F.G.~von~Soden-Fraunhofen,
E.~Mirabella, T.~Peraro, J.~Reichel, M.~Rodgers, J.~Schlenk.}\\
       "CERN"\\
       E-mail: \email{Francesco.Tramontano@cern.ch}}

\abstract{The public code GOSAM for the computation of the one loop virtual corrections
to scattering amplitudes in the Standard Model and beyond is presented. Particular emphasis is
devoted to the interface with other public tools via the Binoth Les Houches Accord.
We show with examples that doing LHC phenomenology including automatically Next to Leading Order
QCD corrections is now handy.}

\FullConference{"Loops and Legs in Quantum Field Theory" 11th DESY Workshop on Elementary Particle Physics \\
		 April 15-20, 2012\\
		 Wernigerode, Germany}

\begin{document}

\section{Introduction}
The recent discovery of a new resonance at the Large Hadron Collider (LHC) at
CERN with mass $125\,GeV$~\cite{exp} is the result of the efforts in the
construction and operation of the accelerator and of the detectors
as well as of the efficiency in the implementation and interpretation of
very detailed analyses on the large amount of informations collected.
Now the target is to perform detailed measurements of all the branching ratios of the
novel particle while searching for further new resonances
in all the accessible mass regions.
The first question is to establish if the new resonance is or not the
Higgs boson.
In this scenario the help of automated tools for the simulation
of signals and backgrounds coming from the Standard Model or beyond
is desirable and is now possible including Quantum Chromo Dynamic (QCD)
corrections up to the Next to Leading Order (NLO)
accuracy in the strong coupling constant.
The QCD at Hadron Colliders play of course the main role
and several fully automated tools exist that include the
leading order matrix element, the leading logarithmic correction
as a parton shower and the subsequent hadronization, leading to full event
simulation~\cite{smc}.
The NLO hard QCD corrections, being in general quite sizable,
also have to be included to get predictions with a theoretical error comparable
with the accuracy attainable at the LHC.
NLO QCD corrections have been successfully matched in full generality
with a parton shower avoiding the double counting that derive by the
fact that the shower produces also some of the emission predicted in the
full NLO calculations~\cite{nlosmc}. In this way a number
of processes have been studied at NLO accuracy plus parton shower.
The deep understanding of fundamental properties and the structure
of the divergences of QCD corrections up to the NLO made the automation
of the subtractions of the divergences and the matching to the parton shower
feasible in full generality. On the other hand progress has been done in the computation
of the virtual part of the NLO correction exploiting the analytic
properties of loop amplitudes. Advanced computer languages and
the increasing cpu performances give now the possibility to automatize
also the computation of the virtual corrections even for complicated hard processes, i.e. 
processes with many legs in the final state. \\
A key ingredient for the full automation of NLO computation is given by the Binoth Les Houches
Accord~\cite{blha} (BLHA) that fixes a set of universal rules for the communication among
different programs that can provide the complementary ingredients needed for the full event simulation. \\
It is the aim of this presentation to illustrate the usage of GOSAM~\cite{gosam},
a new fully automated package for the computation of the one loop
matrix elements, and its interfacing with other programs for the integration
over the phase space of both the real and the virtual corrections to Born processes.

\section{GOSAM}

GOSAM is a python package that has been designed for the generation of codes
for the numerical evaluation of one loop amplitudes.
At the moment the code produced are in Fortran95.
Inside GOSAM a number of libraries are used: the QGRAF~\cite{qgraf} Fortran library
is used for the generation of the Feynman diagrams, while FORM~\cite{form} is used for the algebraic treatment
of the diagrams mainly through the SPINNEY~\cite{spinney} FORM package, the formulas are
then optimized and written by the java package HAGGIES~\cite{haggies}, finally once
a code for a process has been generated it runs calling libraries for the reduction
of tensor integrals and scalar integral evaluation~\cite{integrals}. \\
The construction of the codes has a very simple structure, the aim is to write efficiently
the numerator of each Feynman diagram (or a group of them)
as a multivariate polynomial of the components of the loop momentum.
For each physical phase space point such polynomial in the loop momentum
gets numerical coefficients related to the specific kinematic configuration and
as a tensor integrand sitting on a certain set of Feynman denominators is reduced
and computed numerically by dedicated libraries. \\
After several years since the seminal work
of Passarino and Veltman~\cite{pv}, loop tensor integral reduction techniques have seen new progress.
The original algorithm has been extended to optimize the numerical evaluation
of the form factors especially near unstable configurations~\cite{tensorintegrals}. On the other hand,
the parametric form of the residues of the multipole expansion of the Feynman integrands
has been fully determined and exploited to construct recursive algorithms based on polynomial
interpolation, leading to the
direct numerical determination of the coefficients of the scalar integral~\cite{integrandreduction}.\\
In general two factors determine which method is the fastest or more efficient: the efficiency
of the tensor reduction library and the way the numerator is prepared and matched with the tensor
reduction library. Further, having more then one construction leads to valuable consistency
check on the results. Alternative methods for the construction of the numerator function to match
with loop integrals have also been proposed in~\cite{other}.
A schematic representation of the functionality of the codes generated by GOSAM is given in Figure~1.
\begin{figure}[htb]
\begin{center}
\includegraphics[width=7.5cm]{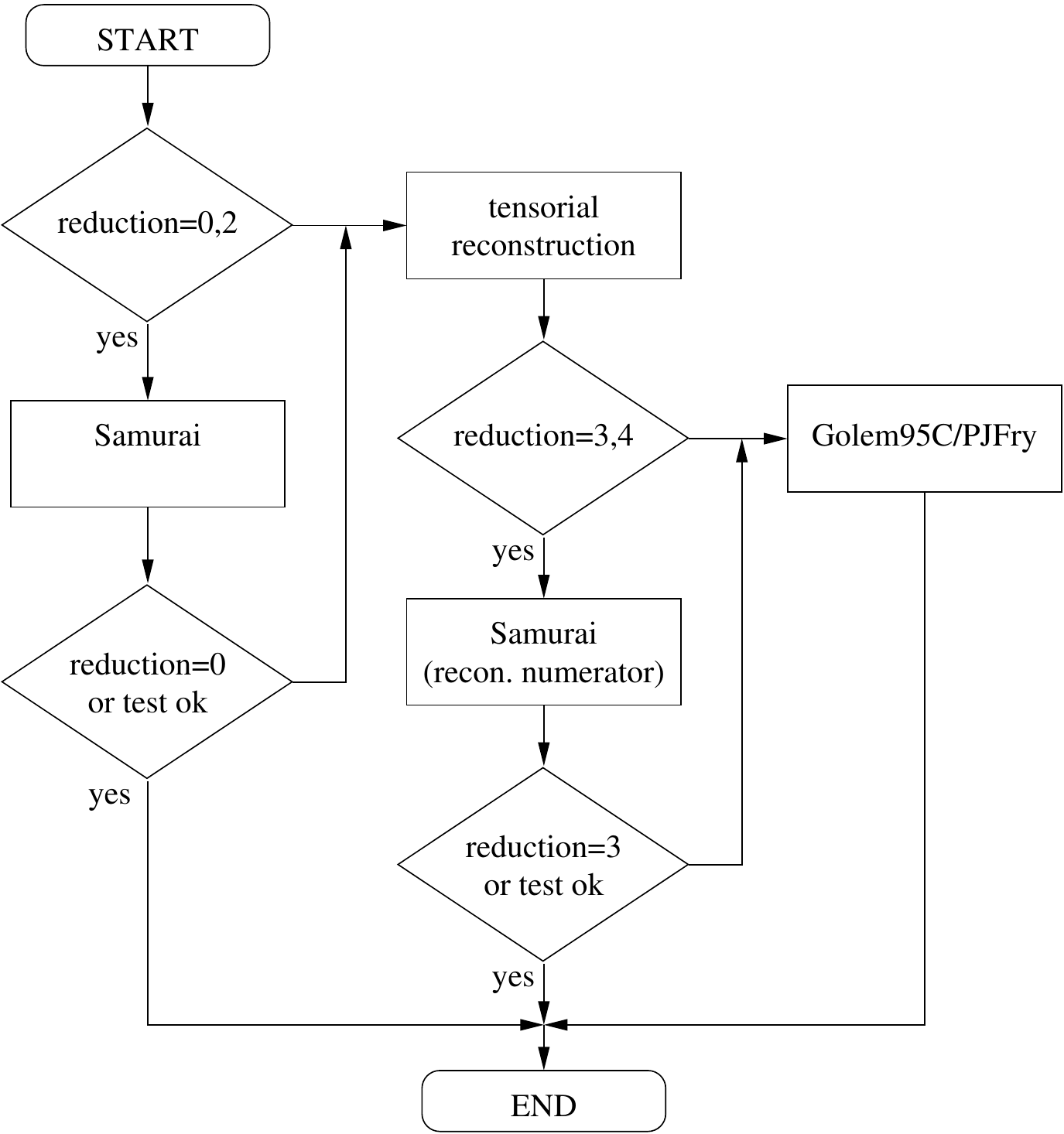} 
\caption{Reduction flow of the GOSAM codes for the evaluation of the numerical virtual amplitudes.}
\label{fig:virt}
\end{center}
\end{figure}
The basic tensor reduction implemented is through the
SAMURAI library, based on the integrand reduction method mentioned above, that also
provides tests to establish the quality of the result. If the test is not passed then the reduction
of the tensor integrals is repeated with GOLEM95. For these cases the codes are equipped with a 
second tensorial representation for the numerator and give more accurate results while remaining
in double precision. The reduction
through SAMURAI is the default because it turns out to be a bit faster then the one through GOLEM95.\\
This switch happens dynamically during the computation of the virtual matrix element,
on a diagram by diagram (or group of diagrams) base.
Finally, for QCD corrections, the renormalised result for the single pole is compared to the expectation
given by the well known formulas expressing the universal behaviour of the infrared QCD divergences,
and if the agreement do not fulfil the required accuracy the whole amplitude is computed again
using GOLEM95.
We have not performed a full study of the correlation among the numerical instabilities of our procedures
and the relative importance of the virtual correction. Experimentally we find that fixing the threshold
for the mismatch of the single pole as
\begin{equation}
\epsilon=\frac{Single~pole(Exact)~~~ - ~~~Single~pole(Numerical)}{\frac{\alpha_s}{2\,\pi}\,Born}
\, \lesssim \, 10^{-4}
\end{equation}
has two consequences: first we note that only a very small fraction of points turns out to be above
threshold (typically one per million for two to three processes) and, second, the results for the
distributions turn out to be in excellent agreement with the one obtained with codes containing
analytic formulas for the virtual amplitudes like the MCFM~\cite{mcfm} suite of programs.
Finally, if this higher level instability trigger finds a phase space point to be still unstable this can be saved
on a file to be reprocessed with higher precision with the same GOSAM code that can be compiled
even with intermediate or quadruple numerical precision.\\
A number of pointwise comparisons has been performed with results obtained mainly with other codes
or by coding amplitudes reported into the appendix of the relative papers, or simply comparing
the numerical results reported for benchmark phase space points.
A sample of the comparisons is given in Table~1, while in Table~2 a list of timings is reported
for the generation of the codes and the relative running time per phase space point.\\
\begin{table}
\caption{Sample of processes for which GOSAM has been compared to the literature.\label{Table:compare}}
\begin{center}
\begin{small}
\begin{tabular}{|ll|}
\hline
\bf process&\bf process\\
\hline
$e^+e^-\to u\overline{u}$          	&$pp\to W^\pm\,jj$ \\
$e^+e^-\to t\overline{t}$             	&$pp\to W^\pm\,b\overline{b}$ (massive b's) \\
$u\overline{u}\to d\overline{d}$&$e^+e^-\to e^+e^-\gamma$ (QED) \\
$g g \to gg$				&$pp \to t\overline{t} H$ \\
$g g \to gZ$				&$pp \to t\overline{t} Z$ \\
$p p \to t\overline{t}$			&$\gamma \gamma \to \gamma \gamma \gamma \gamma $ (fermion loop)  \\
$b g \to H\,b$				&$pp\to W^+W^+jj$ \\
$\gamma \gamma \to \gamma \gamma $ (W loop)&$pp\to b\overline{b} b\overline{b}$ \\
$pp\to W^\pm\,j$ (QCD corr.)	&$pp\to W^+W^- b\overline{b}$ \\
$pp\to W^\pm\,j$ (EW corr.)	&$pp \to t\overline{t}b\overline{b}$ \\
$pp\to W^\pm\,t$ 			&$u\overline{d} \to W^+ ggg$\\
\hline
\end{tabular}
\end{small}
\end{center}
\end{table}
\begin{table}[h]
\begin{center}
{\small \begin{tabular}{|l|r|r|}
\hline
Process                                          &
\multicolumn{1}{|c|}{Generation [s]} &
\multicolumn{1}{|c|}{Evaluation [ms]} \\
\hline
$bg\to Hb$                                       &      236&  2.49\\
$d\bar{d}\to t\bar{t}$            &      324&  4.05\\
$dg\to dg$                                       &      398&  3.08\\
$e^+e^-\to t\bar{t}$\,(LanHEP)            &      180&  1.27\\
$e^+e^-\to u\bar{u}$\,(AutoTools)         &      173&  0.64\\
$gg\to gg$\,(LanHep)                      &     1022&  1.70\\
$gg\to gZ$                                       &      529&  15.18\\
$gg\to t\bar{t}$\,(UFO)                   &     1225&  29.45\\
$H\to\gamma\gamma$                               &      140&  0.24\\
$gb\to e^-\bar{\nu}_et$                          &      337&  2.89\\
$u\bar{d}\to e^-\bar{\nu}_e$                     &       71&  0.09\\
$u\bar{d}\to e^-\bar{\nu}_eg$                    &      154&  1.15\\
$u\bar{u}\to d\bar{d}$                           &      186&  2.06\\
$\bar{u}d\to W^+W^+\bar{c}s$                     &     1295&  17.37\\
$\gamma\gamma\to\gamma\gamma$                    &      597&  6.08\\
\hline
\end{tabular} } \label{timecode}
\caption{Time required for code generation and calculation of one phase-space point. 
The results were obtained with an Intel(R) Core(TM) i7 CPU 950  @ 3.07GHz.
The time for the evaluation of a phase space point is taken as the average
of the time obtained from the evaluation of 100 random points,
where the code was compiled using \texttt{gfortran}
without any optimisation options. The generation of the $R_2$ term was
set to \texttt{explicit}.
}
\end{center}
\end{table}
We generated codes for the NLO QCD corrections for the production at hadron colliders
of four bottom quarks~\cite{fourb} and for the associated production of two $W$ bosons with opposite
charge plus two jets~\cite{wwjj}. Both computations allowed to test the packages on amplitudes
consisting of roughly one thousand diagrams. The running time per phase space point
for such matrix elements is around one second for a single subprocess preventing
the direct integration on a single core of the full virtual matrix elements.
In these cases we proceeded through the re-weighting of
a previously obtained sample of Born level un-weighted events to obtain a reasonable
estimation of the virtual correction. The integration of the real part and of the subtractions
has been performed with the packages MadGraph~\cite{madg}, MadDipole~\cite{madd} and MadEvent~\cite{made}.
The gluon initiated four bottom quarks production virtual amplitudes were integrated for the first time using
for their numerical evaluation a GOSAM code, while the main contributions for the process
$pp \to W^+W^-+2jets$ were already be computed in~\cite{mmrz}. In~\cite{wwjj} we included the
computation of the closed fermion loops with electroweak gauge bosons attached and the
non top resonant contribution form the third generation.\\

\section{Interfacing GOSAM}

Pointwise tests and integrations based on the re-weighting procedure for complicated two to
four one loop processes show how our public package GOSAM can be efficiently used as a fully automatic
generator for the virtual amplitudes. However, that is only a single piece in the task of the full
automatic NLO code generation.
The standard communication setup established in the BLHA
gives then the possibility to link GOSAM to any other program that supports the same standard and 
supplement the other
ingredients. The level of the automation in the generation
of the tree level Born and real corrections, of the subtractions of the divergences as well as in the management
of the integrations and the analysis of the results is already very high.
We exploited the standard to link GOSAM to the SHERPA package that since the release
1.4 is equipped with all the ingredients needed for the parton level next to leading
order simulation of QCD correction to the Standard Model processes, so that the whole NLO code
is generated by a single command-line executable~\cite{proc}.
In this way the generation of the whole code is fully automatic so that once GOSAM
provides the virtual amplitudes all the features already implemented in SHERPA
for the leading order analyses become available for the NLO computation. Representative examples
of distributions obtained with code generated fully automatically
with GOSAM+SHERPA are given in Figures~2, 3 and 4 for the processes $pp \to W^-b\overline{b}$
with massive bottom quarks,
$pp \to W^-+2jets$ and $pp \to W^+W^++2jets$ respectively.
The comparisons have beed done with MCFM~\cite{mcfm} for the former two processes and
with~\cite{mmrz2} for the latter.
\begin{figure}[h]
\includegraphics[width=16pc]{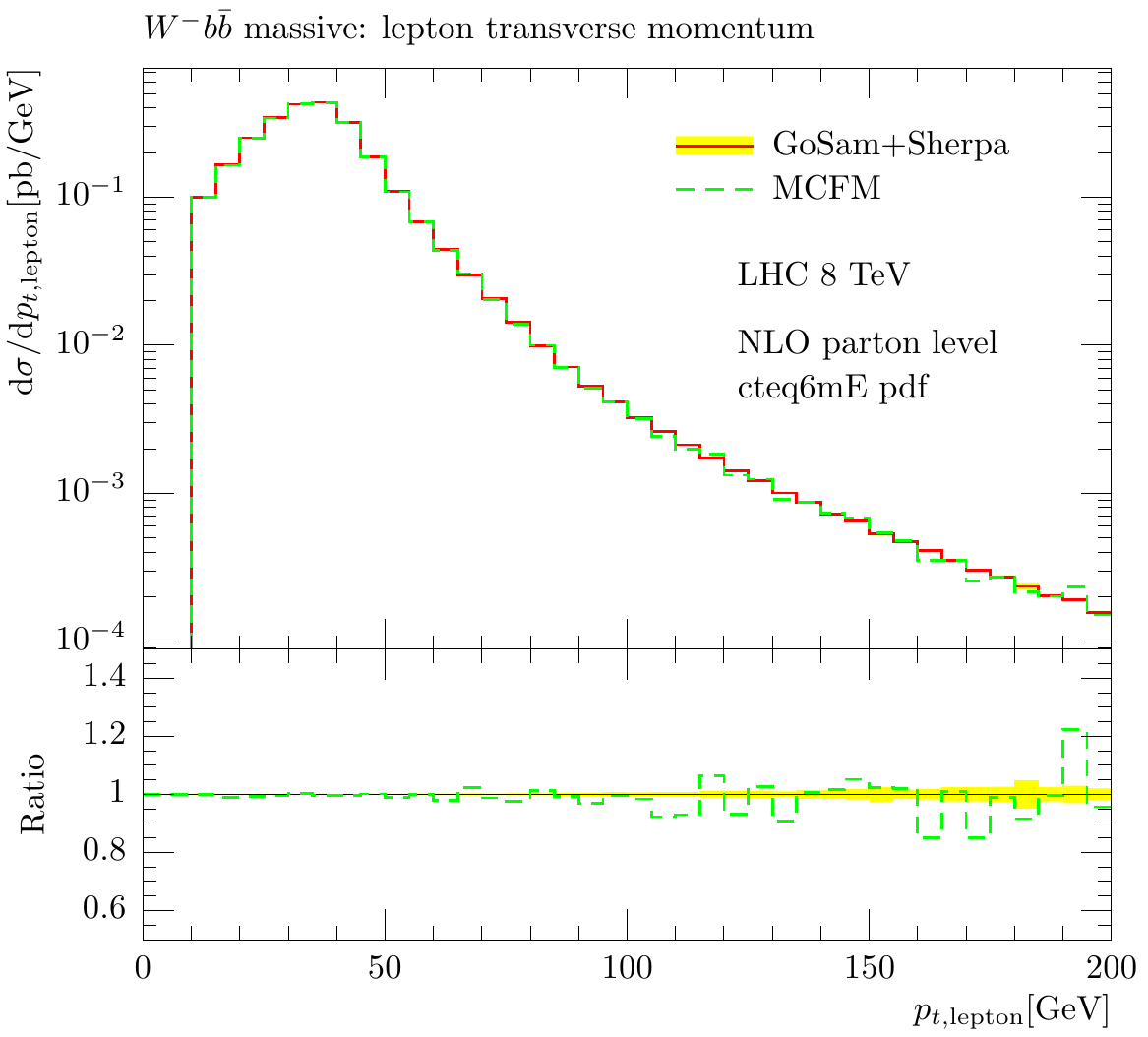}
\hspace{2pc}
\includegraphics[width=16pc]{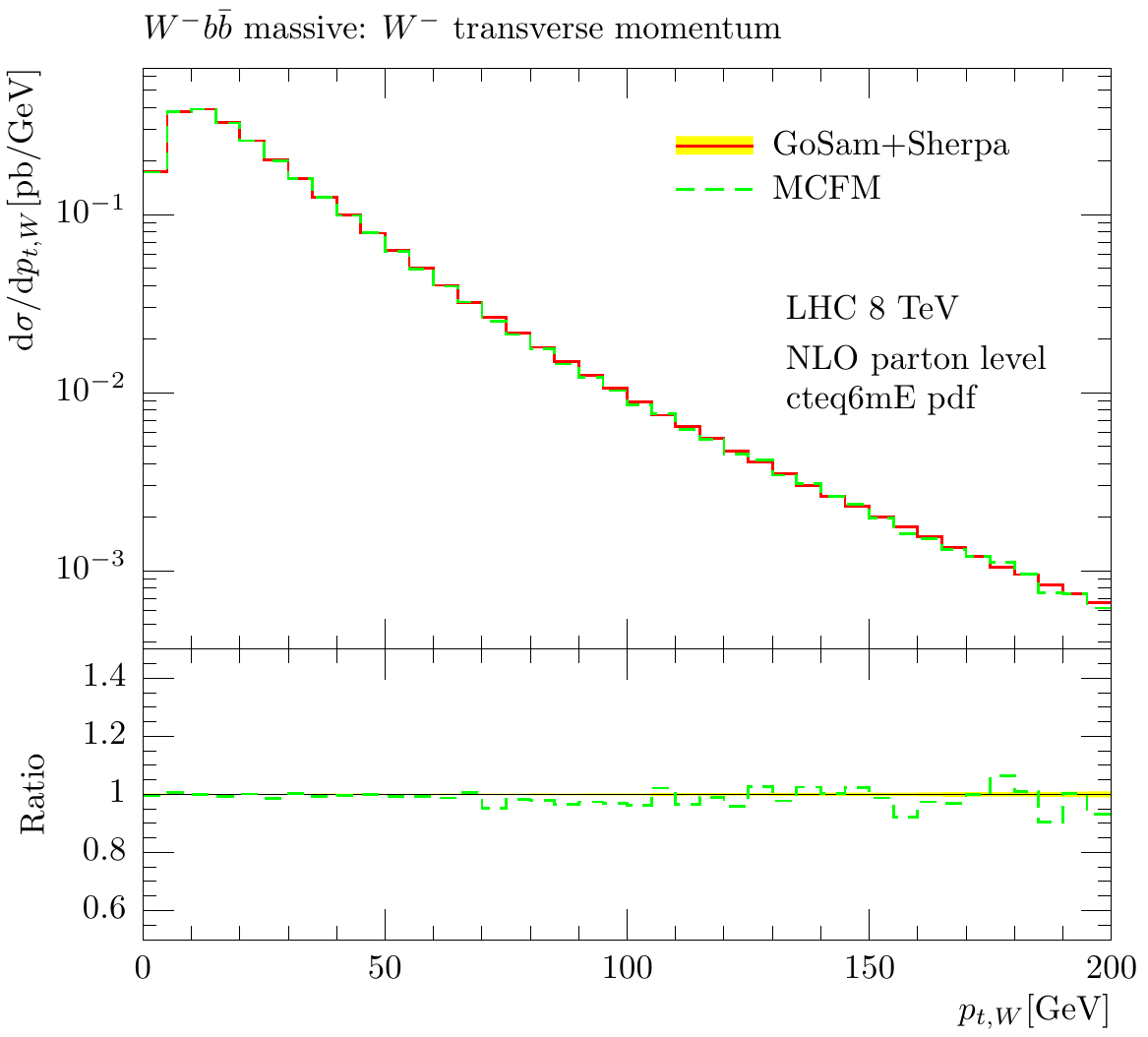}
\caption{Charged lepton and $W^-$ transverse momentum distribution for the
process $pp \to W^-\,b \overline{b}$. Prediction obtained with
GOSAM+SHERPA are compared to MCFM for a set of standard cuts.}
\end{figure}
\begin{figure}[h]
\includegraphics[width=16pc]{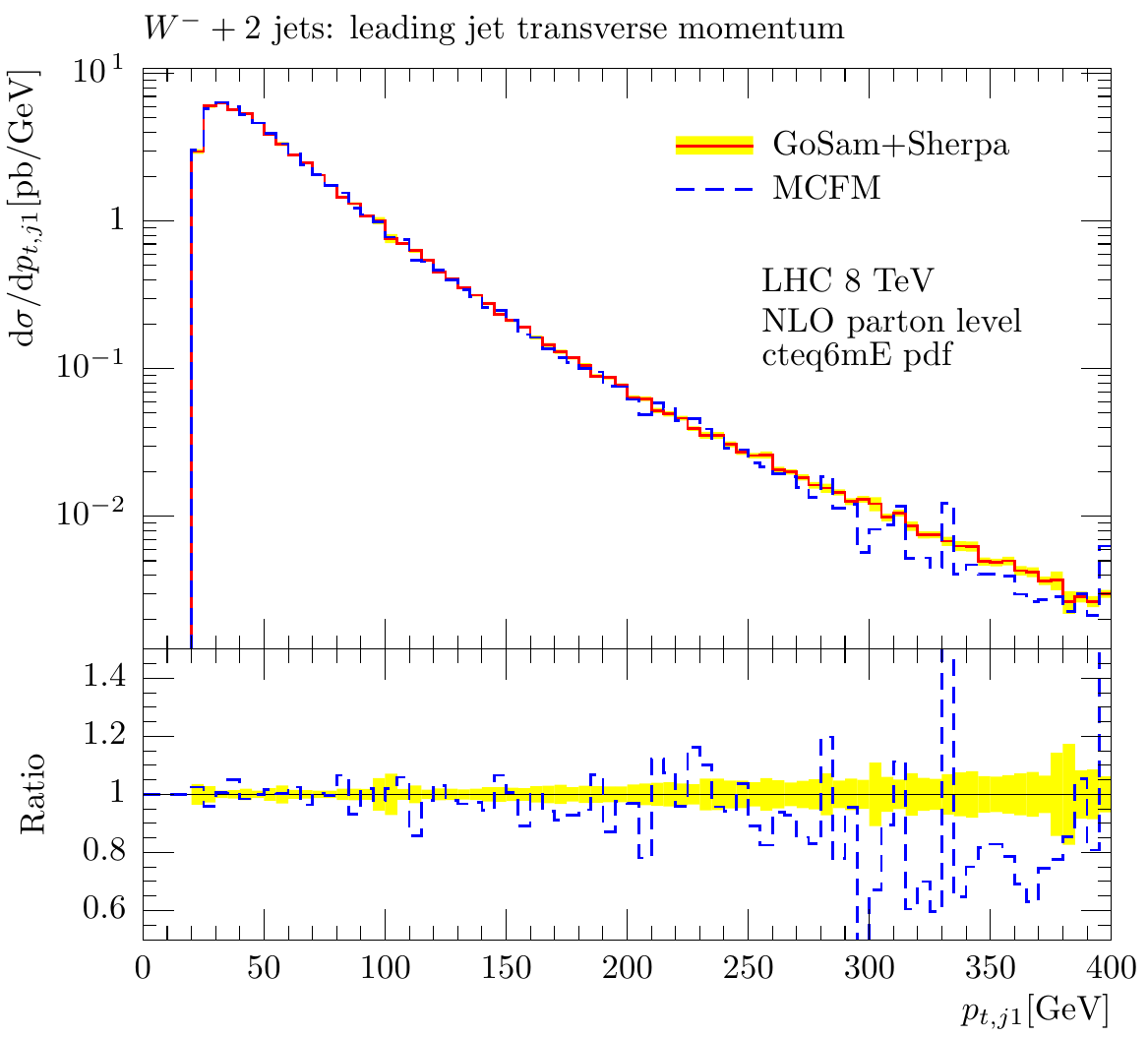}
\hspace{2pc}
\includegraphics[width=16pc]{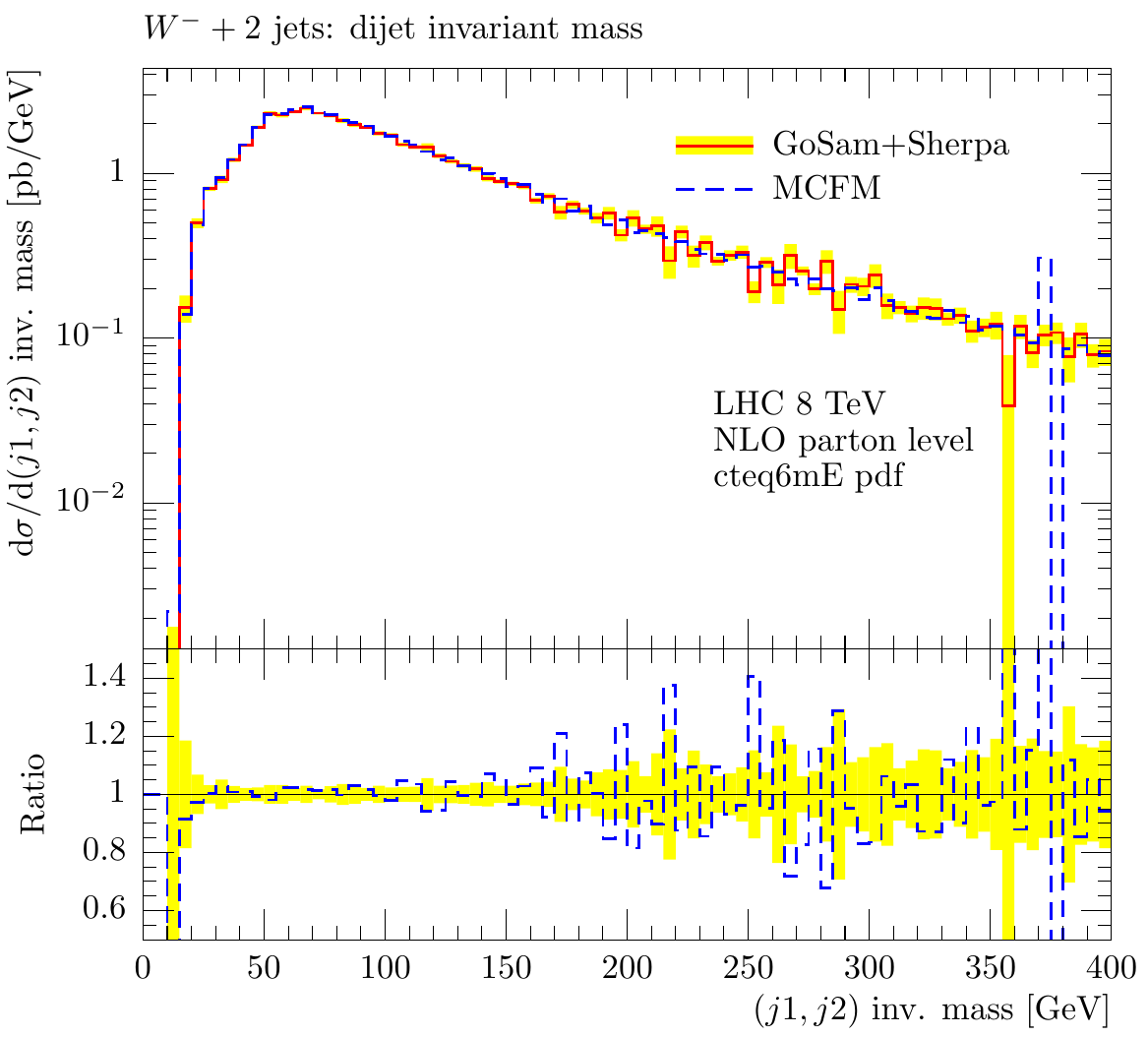}
\caption{Leading jet transverse momentum and diet invariant mass distributions for the
process $pp \to W^-\, + 2jets$. Prediction obtained with
GOSAM+SHERPA are compared to MCFM for a set of standard cuts.}
\end{figure}
\begin{figure}[h]
\includegraphics[width=16pc]{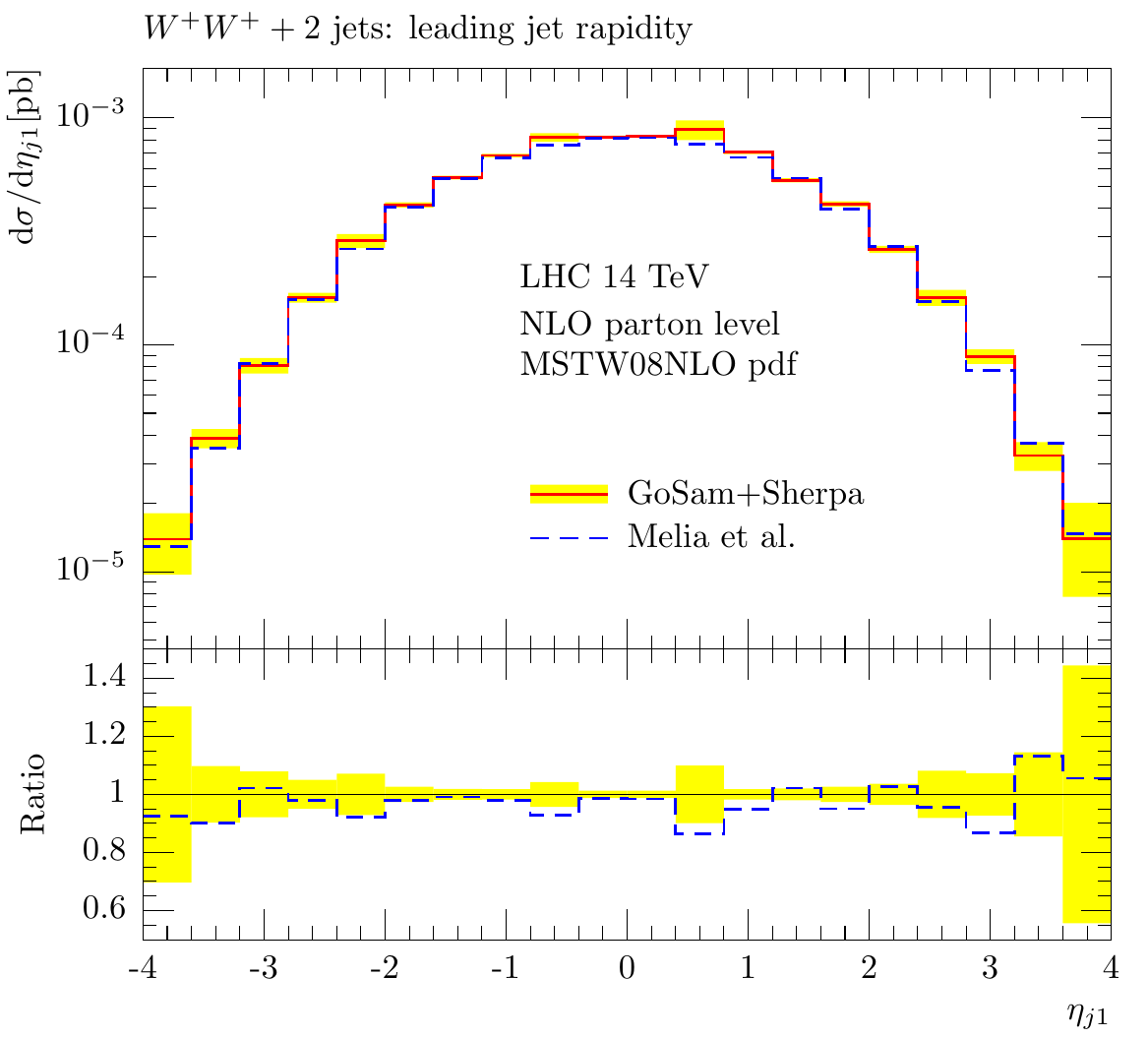}
\hspace{2pc}
\includegraphics[width=16pc]{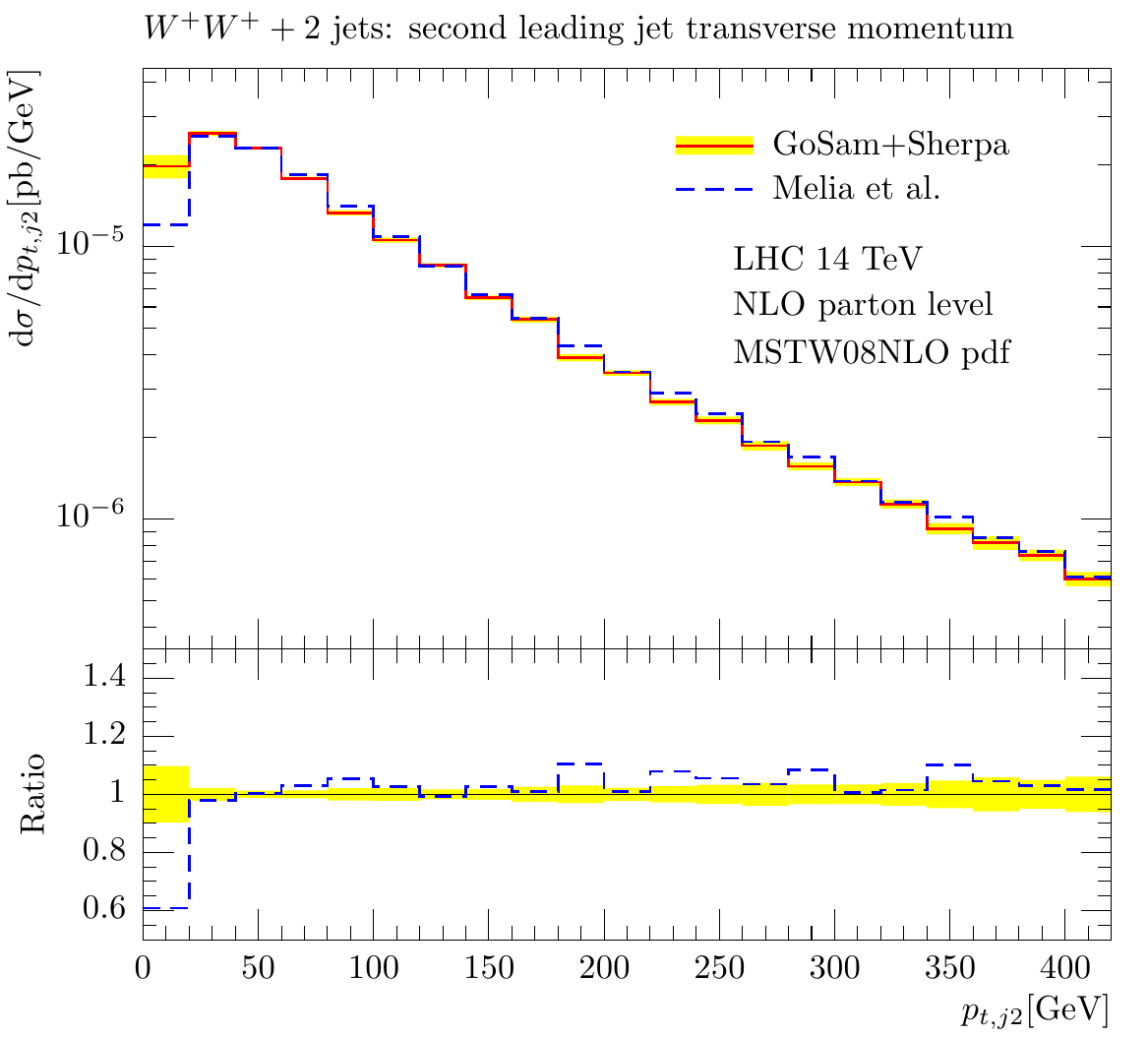}
\caption{Rapidity of the leading jet and second jet transverse momentum distributions for the
process $pp \to W^+\,W^+\, + 2jets$. Prediction obtained with
GOSAM+SHERPA are compared with plots extracted from~\cite{mmrz2} following the
phenomenological analysis reported there.}
\end{figure}
In conclusion we have presented the GOSAM package for the generation
of numerical codes for the virtual amplitudes at one loop and shown how the interface
with other public codes can be exploited to have the fully automatic generation
of programs for the computation of the NLO correction to the hard scattering processes.

\end{document}